\documentclass[prb,twocolumn,groupeaddress]{revtex4-1}

\usepackage{amsmath}
\usepackage{amssymb}
\usepackage{xspace}
\usepackage{graphicx}
\usepackage{grffile}
\usepackage{nicefrac}
\usepackage{color}

\graphicspath{{figs/}}

\begin{document}

\title{Late transition-metal oxides with infinite-layer structure: Nickelates versus cuprates}
\author{Frank Lechermann}
\affiliation{I. Institut f{\"u}r Theoretische Physik, Universit{\"a}t Hamburg, 
Jungiusstr. 9, 20355 Hamburg, Germany}

\begin{abstract}
The correlated electronic structure of the infinite-layer compounds NdNiO$_2$ and
SrCuO$_2$ at stoichiometry and with finite hole doping is compared. Key differences
are elucidated from an advanced first-principles many-body
perspective. Contrary to the charge-transfer insulating cuprate, the self-doped nickelate
remains non-insulating even for large interaction strength, though the 
Ni-$d_{x^2-y^2}$ spectral weight is also gapped in that limit. Hybridization between 
Ni$(3d)$ and Nd$(5d)$ is crucial for the appearance of the self-doping band.
Upon realistic hole doping, Sr$_{1-y}$CuO$_2$ shows the expected mixed 
oxygen-Cu-$d_{x^2-y^2}$ (Zhang-Rice) states at low-energy. In the case of 
Nd$_{1-x}$Sr$_x$NiO$_2$, the self-doping band is shifted to higher energies and a 
doping-dependent $d_{z^2}$-versus-$d_{x^2-y^2}$ competition on Ni is revealed. The 
absence of prominent Zhang-Rice physics in infinite-layer nickelates might be relevant 
to understand the notable difference in the superconducting $T_{\rm c}$'s.

\end{abstract}

\maketitle

\textit{Introduction.---} 
The competing energy scales of charge-transfer and on-site Coulomb kind
are important for various properties of transition-metal (TM) 
oxides~\cite{zaa85}, including high-temperature superconductivity in 
doped layered cuprates~\cite{bed86}. Recently, Li {\sl et al.} reported 
superconductivity up to $T_{\rm c}=15\,$K in the infinite-layer (IL) 
nickelate NiNdO$_2$ with hole doping~\cite{li19}. It was achieved by thin-film
generation via soft-chemistry topotactic reduction on a SrTiO$_3$ substrate.
Albeit these results are debated in view of the underlying structural 
details~\cite{li219,zho19}, this first successful finding of a respectable 
superconducting regime in conjunction with a layered-cuprate analogon from the 
nickelate family is 
remarkable~\cite{gu19,hep20,hu19,hir19,jia19,jiang19,nom19,rye19,si19,wer20,wu19,zha20}.

The IL architecture usually refers to perovskite(-like) ${\cal A}$BO$_n$ crystals in 
which the apical oxygens are missing, hence BO$_2$ square lattices are stacked
with separating ${\cal A}$ layers. Transition-metal oxides of IL kind are e.g. known 
for iron~\cite{tsu07}, nickel~\cite{cre83,hay03} and copper~\cite{sie88,smi91,azu92} 
compounds. Generally for late transition-metal oxides, the TM$(3d)$ subshell of 
$t_{2g}=\{d_{xz},d_{yz},d_{xy}\}$ character is completely filled and orbitals from 
the $e_g=\{d_{z^2}, d_{x^2-y^2}\}$ sector are partially filled. Whereas most cuprates 
are prototypical charge-transfer insulators with a comparatively small charge gap, many 
nickelates also carry substantial Mott-Hubbard character. Furthermore, while IL 
cuprates host the common formal Cu$^{2+}$ oxidation state, as the perovskite compounds, 
IL nickelates formally host the uncommon Ni$^{+}$ oxidation state~\cite{lee04}, when 
Ni$^{2+}$ is usually realized in other nickel oxides. Thus one expects crucial 
normal-state differences between IL cuprates and nickelates at stoichiometry and with 
finite doping, largely effecting also the superconducting instability.
\begin{figure}[t]
\includegraphics*[width=8.5cm]{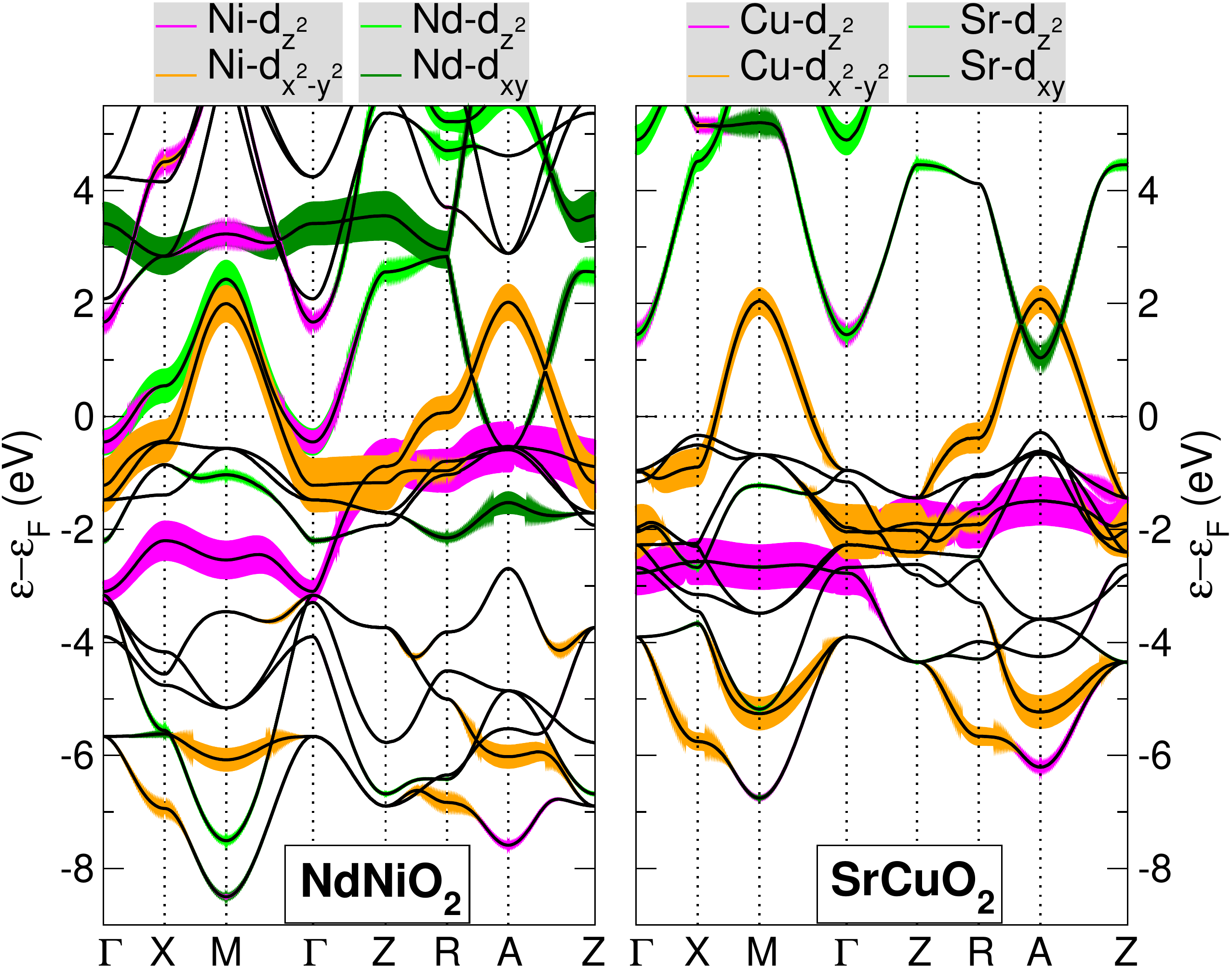}\\[-0.1cm]
\caption{(color online) DFT band structure of NdNiO$_2$ (left) and
SrCuO$_2$ (right) along high-symmetry lines in the Brillouin zone. 
Color coded are the dominant orbital weights
Ni-$d_{z^2}$, Ni-$d_{x^2-y^2}$ and
Nd-$d_{z^2}$, Nd-$d_{xy}$ as well as
Cu-$d_{z^2}$, Cu-$d_{x^2-y^2}$ and
Sr-$d_{z^2}$, Sr-$d_{xy}$ close to the Fermi level.}
\label{fig:dft}
\end{figure}

In this paper, we focus on a qualitative comparison of the long-known cuprate SrCuO$_2$ 
with the isostructural nickelate NdNiO$_2$. The former compound is a 
charge-transfer insulator~\cite{ter96,*ter93} and becomes a high-temperature superconductor 
upon doping and further alloying~\cite{smi91,azu92}. On the other hand, the measured 
conductivity~\cite{li19} in the stoichiometric IL nickelate does not suggest a sizable 
charge gap. Our realistic many-body study takes care of the subtle interplay between 
charge-transfer and Mott-Hubbard physics in these late TM oxides. Full charge self-consistency 
and full rotational invariance in the local TM$(3d)$ Coulomb interaction is combined
with the inclusion of electron correlations stemming from oxygen sites. Important 
differences between both compounds, and points for NdNiO$_2$ toward a coexistence of 
Mott-critical layers with residual metallicity due to a subtle coupling between NiO$_2$ 
and Nd sheets. Upon hole doping, the stronger multi-orbital Mott-Hubbard character of the 
nickelate at low energy becomes evident, different from the Zhang-Rice physics of the 
effective one-orbital cuprate. 
\begin{figure*}[t]
\raggedright\includegraphics*[height=4.65cm]{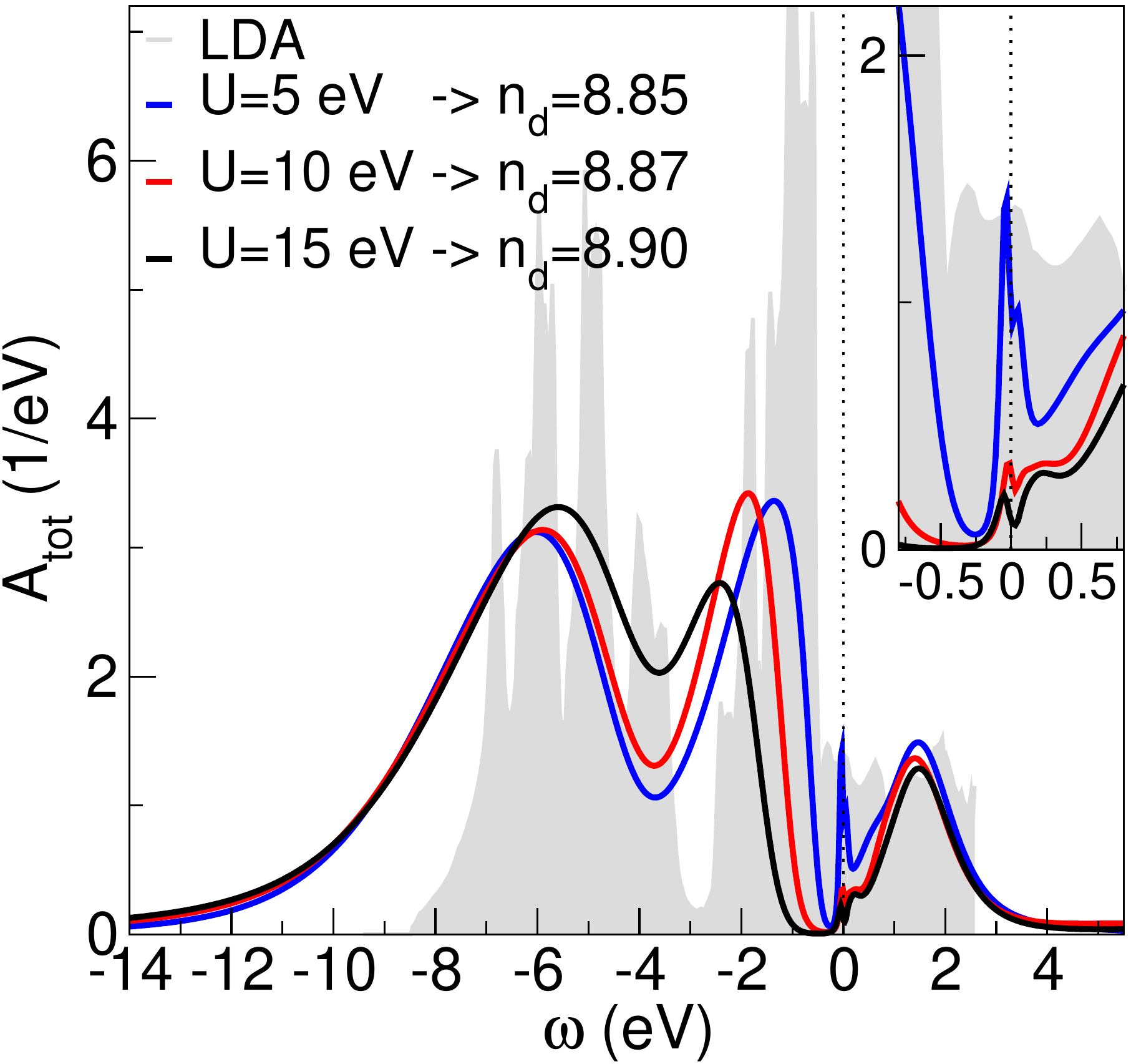}\hspace*{0.3cm}
\includegraphics*[height=4.65cm]{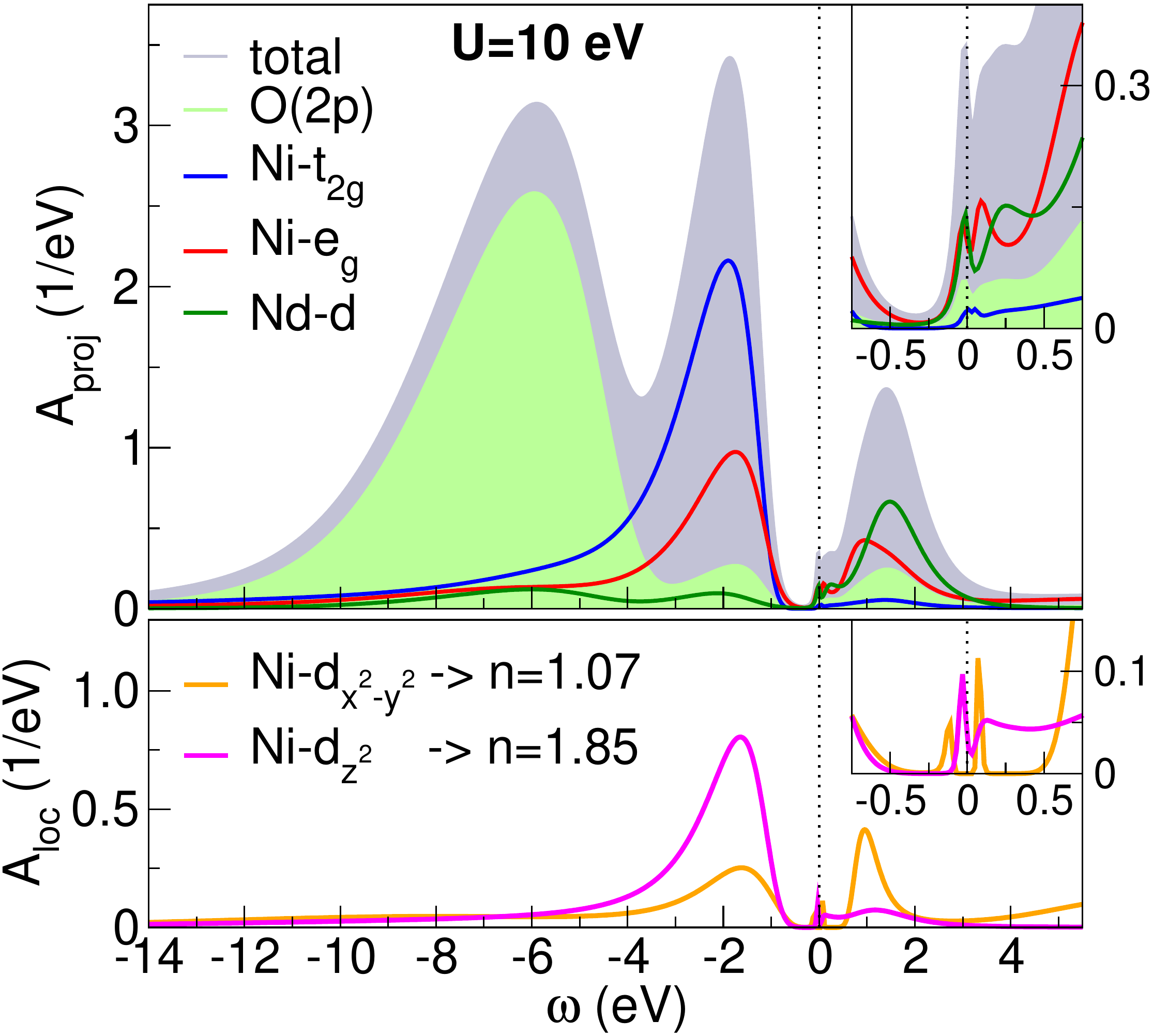}\hspace*{0.1cm}
\includegraphics*[height=4.65cm]{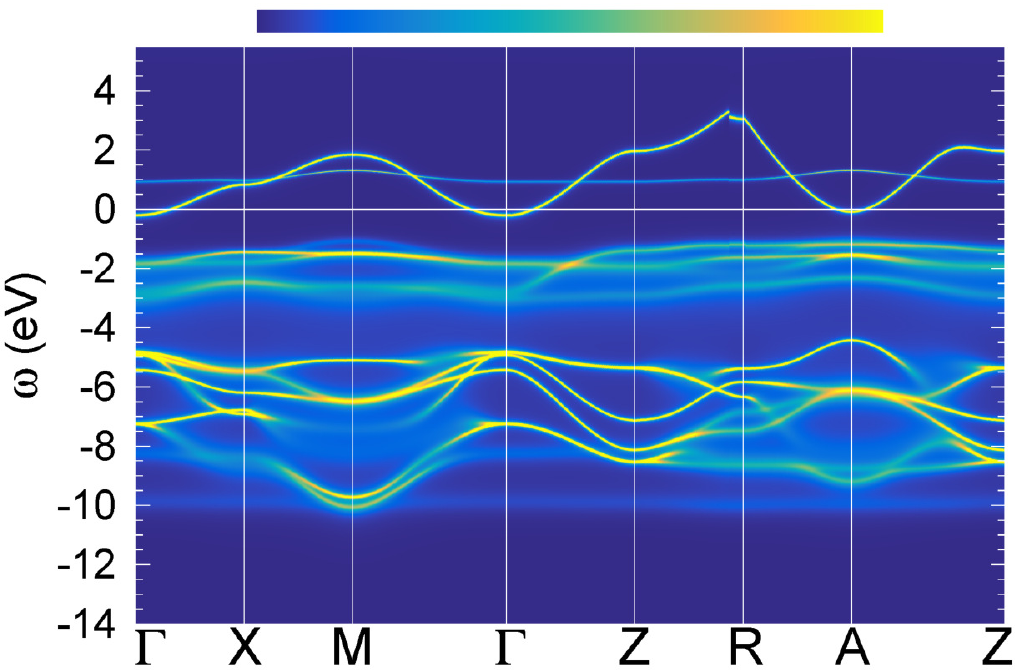}\\[-0.2cm]
\raggedright(a)\hspace*{5cm}(b)\hspace*{5cm}(c)\\[0.2cm]
\raggedright\includegraphics*[height=4.65cm]{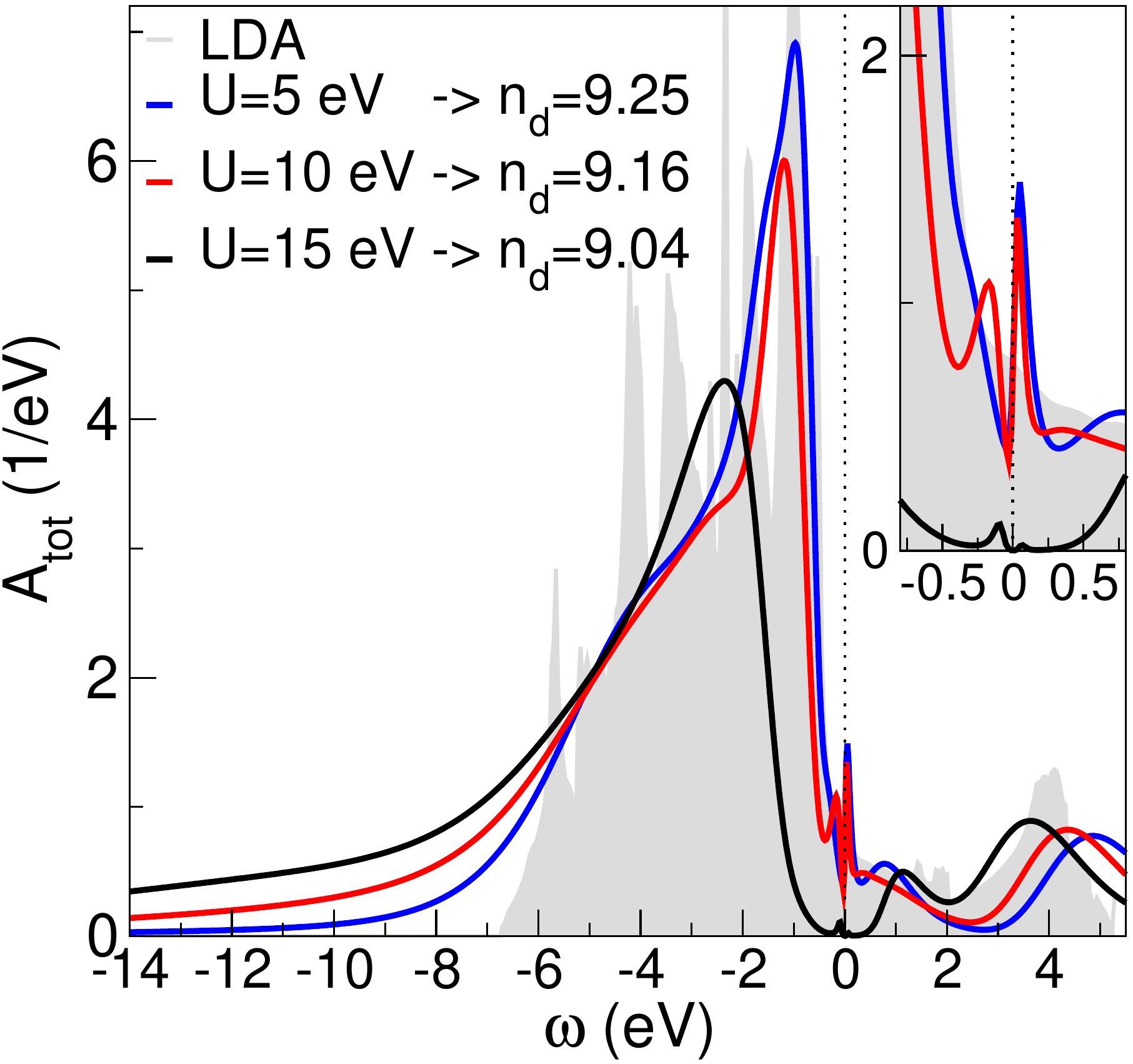}\hspace*{0.3cm}
\includegraphics*[height=4.65cm]{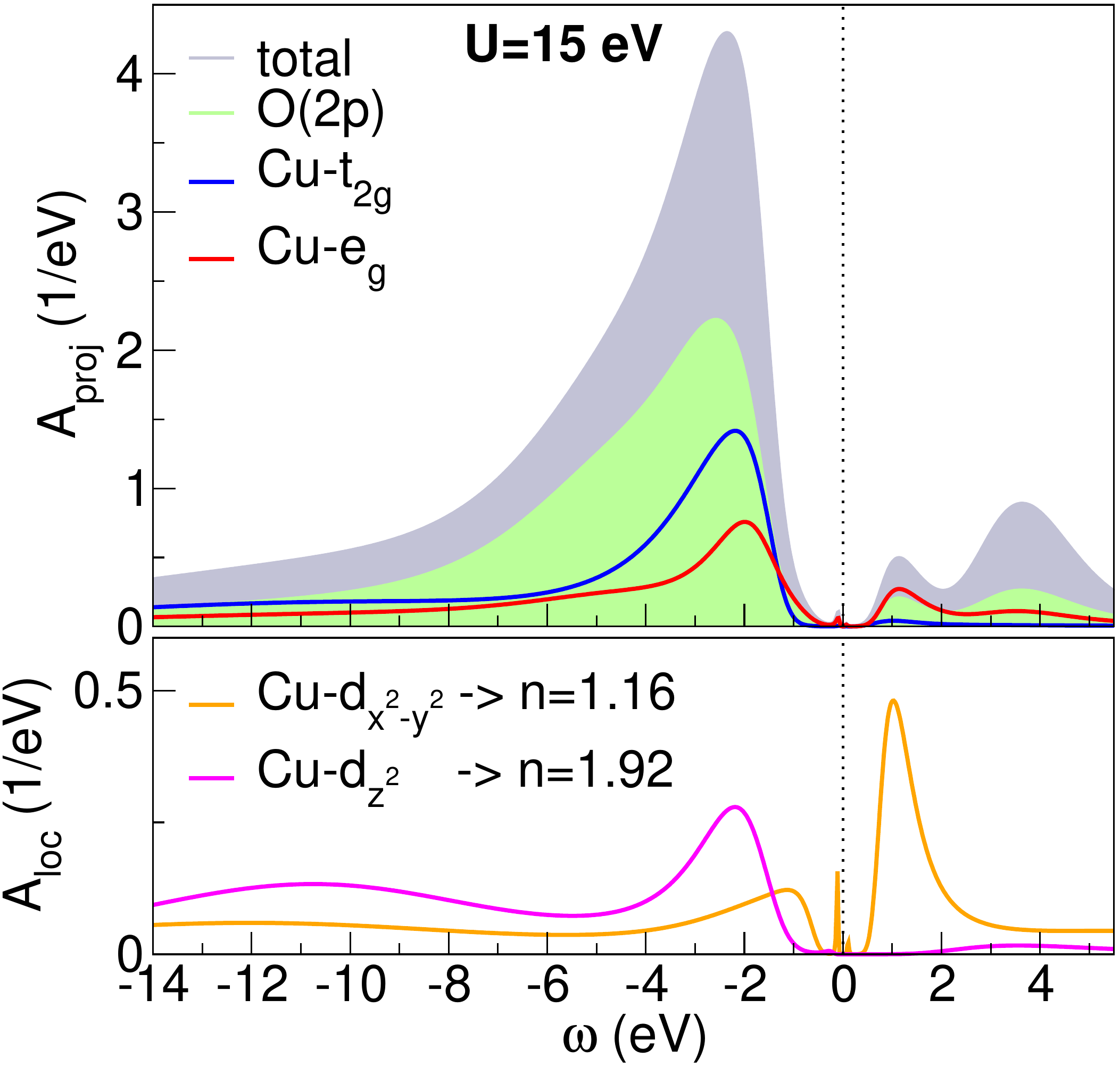}\hspace*{0.4cm}
\includegraphics*[height=4.65cm]{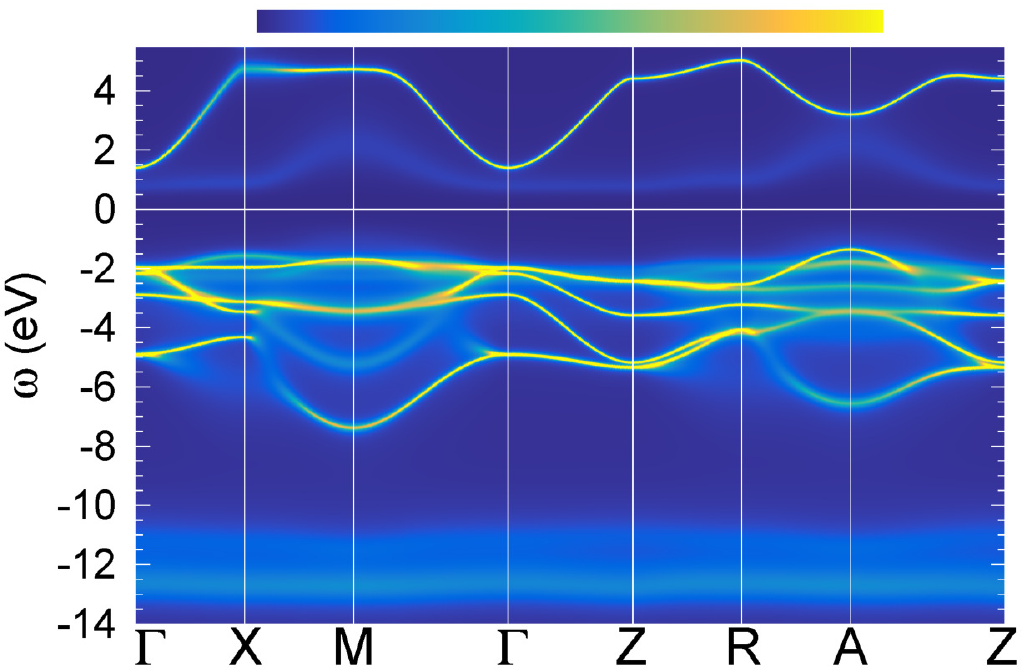}\\[-0.2cm]
\raggedright(d)\hspace*{5cm}(e)\hspace*{5cm}(f)
\caption{(color online) DFT+sicDMFT spectral data of NdNiO$_2$ (top,a-c) and SrCuO$_2$
(bottom,d-f).
(a,d) Total spectral function for $U=5, 10, 15$\,eV (insets: low-energy blow up) joint
with the TM$(3d)$ occupation $n_d$.
(b,e) orbital- and site-projected spectral functions (upper part), and local TM-$e_g$
spectral functions (lower part) joint with respective fillings $n$.
(c,f) {\bf k}-resolved spectral function $A({\bf k},\omega)$ along high-symmetry 
lines. (b,c) are based on $U=10$\,eV and (e,f) on $U=15$\,eV.}
\label{fig:pristine}
\end{figure*}

\textit{Theoretical approach.---}
We employ the charge self-consistent combination~\cite{gri12} of density functional 
theory (DFT), self-interaction correction (SIC) and dynamical mean-field theory (DMFT) 
in the DFT+sicDMFT framework~\cite{lec19}. There, TM sites of Ni or Cu 
chemical entity act as DMFT impurities and Coulomb interactions on O enter 
within SIC on the pseudopotential level~\cite{kor10}. The DFT part consists of a 
mixed-basis pseudopotential code~\cite{els90,lec02,mbpp_code} and the SIC is applied 
to the $2s$ and the $2p$ orbitals of oxygen via weight factors $w_p$ (see 
Ref.~\onlinecite{kor10} for more details). While the O$(2s)$ orbital is by default fully 
corrected with $w_p=1.0$, the common choice~\cite{kor10,lec19} $w_p=0.8$ is used for  
O$(2p)$ orbitals. Note that we put the Nd$(4f)$ states in the pseudopotential
core.
Continuous-time quantum Monte Carlo in hybridzation expansion~\cite{wer06} as implemented
in the TRIQS code~\cite{par15,set16} is utilized to solve the DMFT problem. In the DMFT 
correlated subspace, a five-orbital full Slater-Hamiltonian is applied to the TM 
projected-local orbitals~\cite{ama08}. The fully-localized double-counting 
scheme~\cite{ani93} is applied.
All many-body calculations are performed in the paramagnetic regime for the system 
temperature $T=193$\,K. Maximum-entropy and Pad{\'e} methods are employed for the 
analytical continuation from Matsubara space onto the real-frequency axis. 
Stoichiometric lattice parameters are overtaken from experiment~\cite{li19,smi91}. 

\textit{Stoichiometric case.---}
Figure~\ref{fig:dft} displays the nonmagnetic DFT band structures based on the local 
density approximation (LDA) for NdNiO$_2$ and SrCuO$_2$. Generally, the TM$(3d)$ levels 
are closest to the Fermi level $\varepsilon^{\hfill}_{\rm F}$, while the dominant part 
of O$(2p)$ lies deeper in energy. The oxygen $2p$ levels are much more intertwined with
TM$(3d)$ in the cuprate, pointing to a stronger charge-transfer character of
SrCuO$_2$. We estimate the charge-transfer energy from 
$\Delta=\varepsilon_d-\varepsilon_p$, where $\varepsilon_d$ and $\varepsilon_p$
amount to the average energy of the TM$(3d)$ and O$(2p)$ level as computed 
in DFT+SIC~\cite{lec19}. This results in $\Delta_{{\rm NdNiO}_2}=5.0$\,eV and an indeed 
much lower $\Delta_{{\rm SrCuO}_2}=1.3$\,eV. Similar calculations for rocksalt NiO 
resulted in $\Delta_{\rm NiO}=4.5$\,eV~\cite{lec19}.
While SrCuO$_2$ displays a single-sheet Fermi surface of dominant Cu-$d_{x^2-y^2}$ 
character, a second band crosses $\varepsilon^{\hfill}_{\rm F}$ in NdNiO$_2$, giving 
rise to electron pockets at $\Gamma$ and A. 
In both compounds, the Fermi sheets overall enclose a volume corresponding to one 
electron. The additional weakly-filled band in the nickelate has mixed character 
of mainly Nd-$d_{z^2}$ and Ni-$d_{z^2}$ around $\Gamma$ as well as of Nd-$d_{xy}$ and 
Ni-$d_{xz}$, Ni-$d_{yz}$ around A. It marks the self-doped nature of the IL nickelate.
Similar hybridizations around $\Gamma$ and A may also be observed in the cuprate case, 
along with the parallel contributions of Sr-$d_{z^2}$ and Sr-$d_{xy}$. However there, 
the corresponding band above the dominant Cu-$d_{x^2-y^2}$ dispersion does not cross 
the Fermi level. The Ni-$e_g$ based nearest-neighbor hoppings $t$ to 
Nd-$d_{z^2,xy}$, as extracted in a wide-energy picture from the Wannier-like
projected-local-orbital formalism~\cite{ama08}, read 
$t_{{\rm Ni-}d_{z^2}}^{{\rm Nd-}d_{z^2}}=18$\,meV and
$t_{{\rm Ni-}d_{z^2}}^{{\rm Nd-}d_{xy}}=-69$\,meV. Such hoppings between
Ni-$d_{x^2-y^2}$ and Nd-$d_{z^2,xy}$ are zero. The additional low-energy 
relevance of Ni-$d_{z^2}$ from the $e_g$ sector in IL nickelates has already been 
emphasized by Lee and Pickett~\cite{lee04}. Note that the TM$(4s)$ level hybridzes 
over a large energy range (not explicitly displayed), with appreciable weight in the 
O$(2p)$-block bottom, close to $\varepsilon^{\hfill}_{\rm F}$ and well above the 
Fermi level. It does not play a relevant role for the encountered nickelate 
fermiology.

For the realistic many-body description beyond LDA, we take up a pragmatic
position concerning the local Coulomb parameters. In late TM oxides,
a value $J_{\rm H}=1$\,eV is a common choice for the Hund's exchange. For the better-screened
Hubbard $U$, a value of $10$\,eV recently provided very good agreement between theory and
experiment for stoichiometric as well as Li-doped NiO within DFT+sicDMFT~\cite{lec19}.
While some authors expect a smaller $U$ in IL nickelate, the value for cuprates is
usually expected larger than for nickelates. Hence, we start with the three different
values $U=5, 10, 15$\;eV to cover the principally possible interaction space.
\begin{figure}[t]
\raggedright\includegraphics*[width=4.2cm]{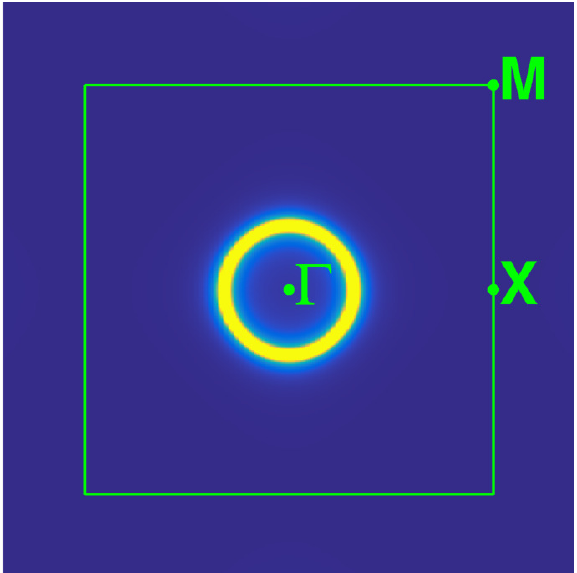}\hspace*{0.1cm}
\includegraphics*[width=4.2cm]{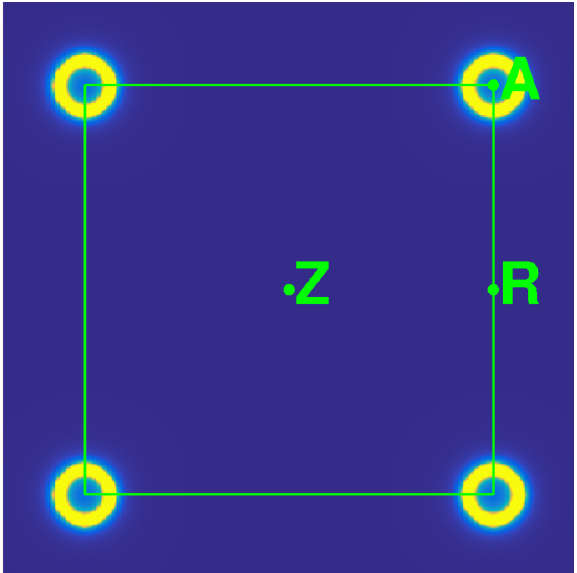}
\caption{(color online) Interacting ($U=10\,$eV) Fermi surface of NdNiO$_2$ in 
the $k_z=0$ (left) and the $k_z=0.5$ (right) plane of the Brillouin zone (green square)
at $T=193$\,K.}
\label{fig:fs}
\end{figure}

Figures~\ref{fig:pristine}a,d show the total ${\bf k}$-integrated spectral function from 
DFT+sicDMFT using the three different $U$ values for NdNiO$_2$ and SrCuO$_2$. 
The O$(2p)$ block is shifted to deeper energies by $\sim 1$\,eV for the former
compound, due to the underestimation of the charge-transfer energy $\Delta$ within LDA. 
As expected, the cuprate compound becomes insulating at large $U$ with a gap of about 
1.5\,eV, i.e. on the order of $\Delta$. For the present largest choice of
$U$, there is still some minor in-gap weight close to zero energy. A large
value of $U>10$\,eV is not unreasonable in extended DMFT schemes for cuprates~\cite{cho16},
yet antiferromagnetic ordering surely additionally supports gap opening.
The nickelate compound avoids an insulating state even for $U=15$\,eV, i.e. the self-doped 
character remains robust up to large interaction strengths. Notably, the TM-$d_{x^2-y^2}$ 
weight is essentially gapped for NdNiO$_2$ at $U=10$\,eV and for SrCuO$_2$ at $U=15$\,eV, 
which from that viewpoint renders the nickelate more strongly correlated. The TM$(3d)$ 
occupation $n_d$ approaches the formal $d^9$ value in both cases with growing $U$. To 
proceed with further details, we narrow down the range of $U$ values to a single one for 
each compound via comparison to experiment: SrCuO$_2$ is a verified charge-transfer 
insulator, thus $U=15$\,eV is chosen; NdNiO$_2$ appears only weakly conducting, therefore 
we discard $U=5$\,eV, and stick to $U=10$\,eV as a well-established value for nickelates. 
For the rest of the paper, these two choices should ensure a qualitatively reliable 
comparison of the correlated electronic structure of both compounds.

The orbital- and site projected spectral functions in Figs.~\ref{fig:pristine}b,e underline
the different $O(2p)$ position in the nickelate and the cuprate. In the case of SrCuO$_2$,
the lower Hubbard band (around $-$12\,eV) lies well below O$(2p)$, whereas for NdNiO$_2$ 
(around $-$9\,eV) it is located in the deeper-energy part of O$(2p)$. Both materials display
sizable charge-transfer character, but with much stronger fingerprint in SrCuO$_2$. The
charge-transfer signature in NdNiO$_2$ is weaker as e.g. in NiO~\cite{lec19}. Both TM-$e_g$
contributions, i.e. $x^2-y^2$ and $z^2$, are gapped for SrCuO$_2$, while the $z^2$ character
takes part in the self-doped state, along with Nd$(5d)$, in the case of the nickelate. 
At $U=10$\,eV, the $x^2-y^2$ orbital in NdNiO$_2$ still marks contributions 
at lower energy, which are finally completely gone for $U=15$\,eV. The $e_g$ occupations
show that Ni-$d_{z^2}$ is indeed further away from complete filling than Cu-$d_{z^2}$, and
hence more susceptible to charge fluctuations. Thus, a novel variant of orbital selectivity 
emerges, with localized Ni-$d_{x^2-y^2}$ and weakly-itinerant Ni-$d_{z^2}$ thanks to 
hybridization with Nd$(5d)$ in a self-doped manner. Though coupling between localized and 
itinerant electrons via e.g. inter-orbital effects on Ni is evident,  a straightforward 
Kondo picture in terms of Nd-$d_{z^2,xy}$ allying with localized Ni-$d_{x^2-y^2}$ does 
seemingly not apply.
The {\bf k}-resolved spectral functions in Figs.~\ref{fig:pristine}c,f confirm the 
previous statements on cuprate gap, nickelate self doping, and TM$(3d)$ vs. O$(2p)$ 
position. Furthermore, in the NdNiO$_2$ case the $dd$-coupling of Ni-$e_g$ to Nd 
is favored for the self-doping band: compared to the LDA result which locates the 
electron pocket at A deeper in energy, the many-body calculation intensifies the stronger 
Ni-$d_{z^2}$-to-Nd-$d_{z^2}$ hybridized electron pocket at $\Gamma$. Finally, the
plotted nickelate Fermi surface in Fig.~\ref{fig:fs} displays circular sheets 
around $\Gamma$ and A corresponding to the electron pockets of the self-doping band.
Note again that the spectral weight of the hole band of mainly Ni-$d_{x^2-y^2}$ character 
is fully transferred to Hubbard bands at the given interaction strength due to 
the specific orbital-selective Mott transition in NdNiO$_2$.

\textit{Hole-doped case.---}
\begin{figure*}[t]
\raggedright\includegraphics*[height=5.4cm]{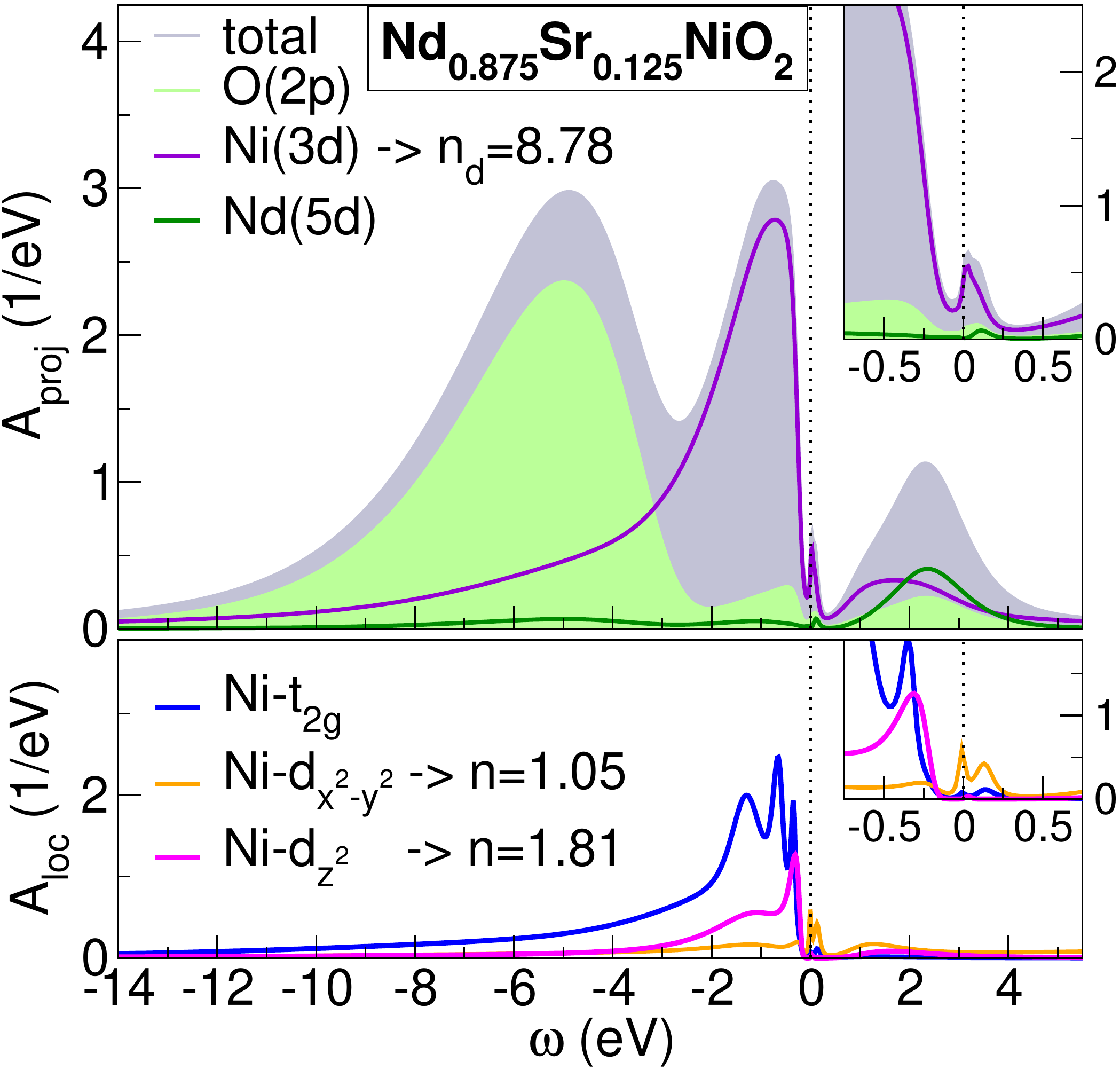}\hspace*{0.3cm}
\includegraphics*[height=5.4cm]{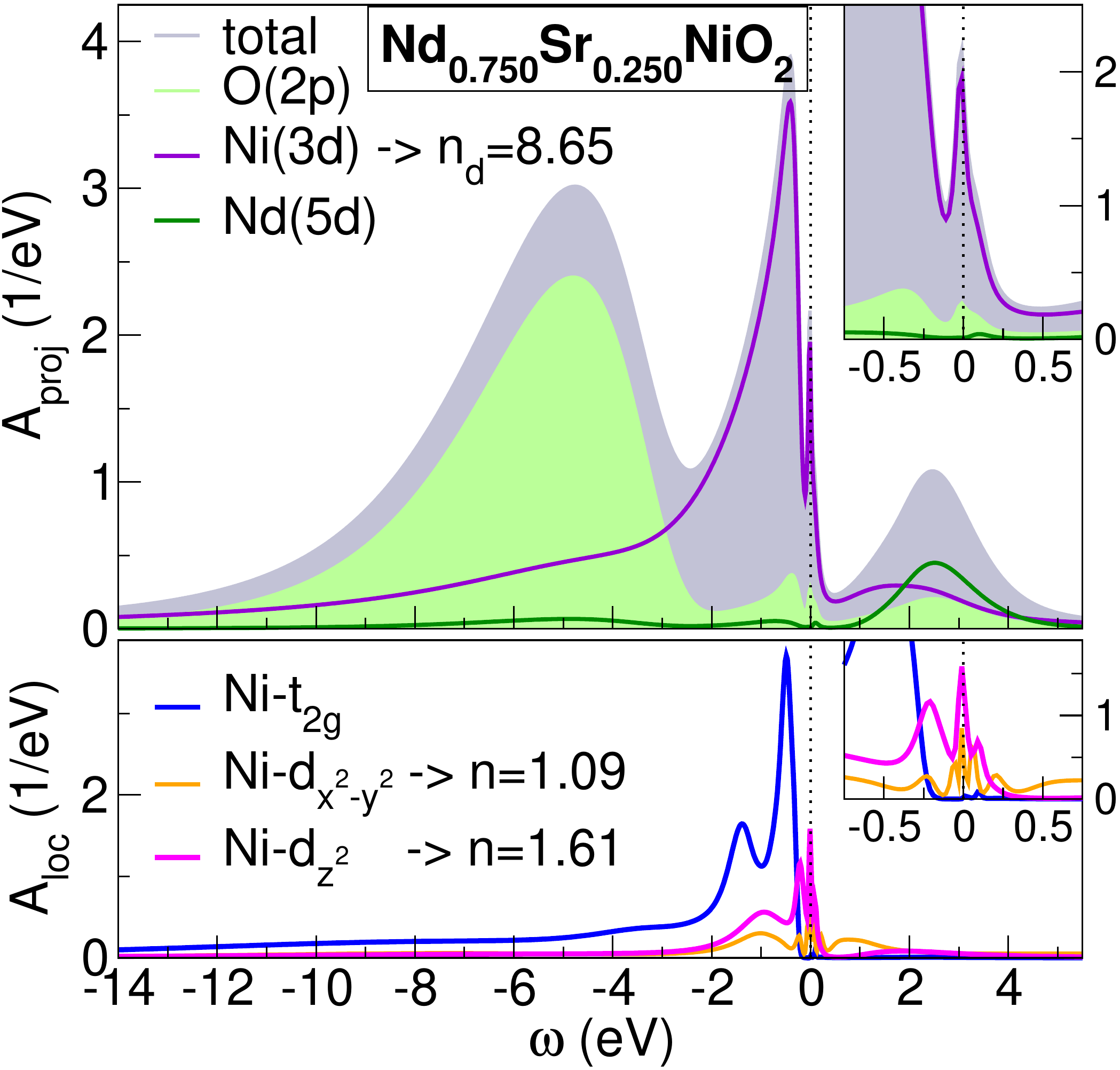}\hspace*{0.3cm}
\includegraphics*[height=5.4cm]{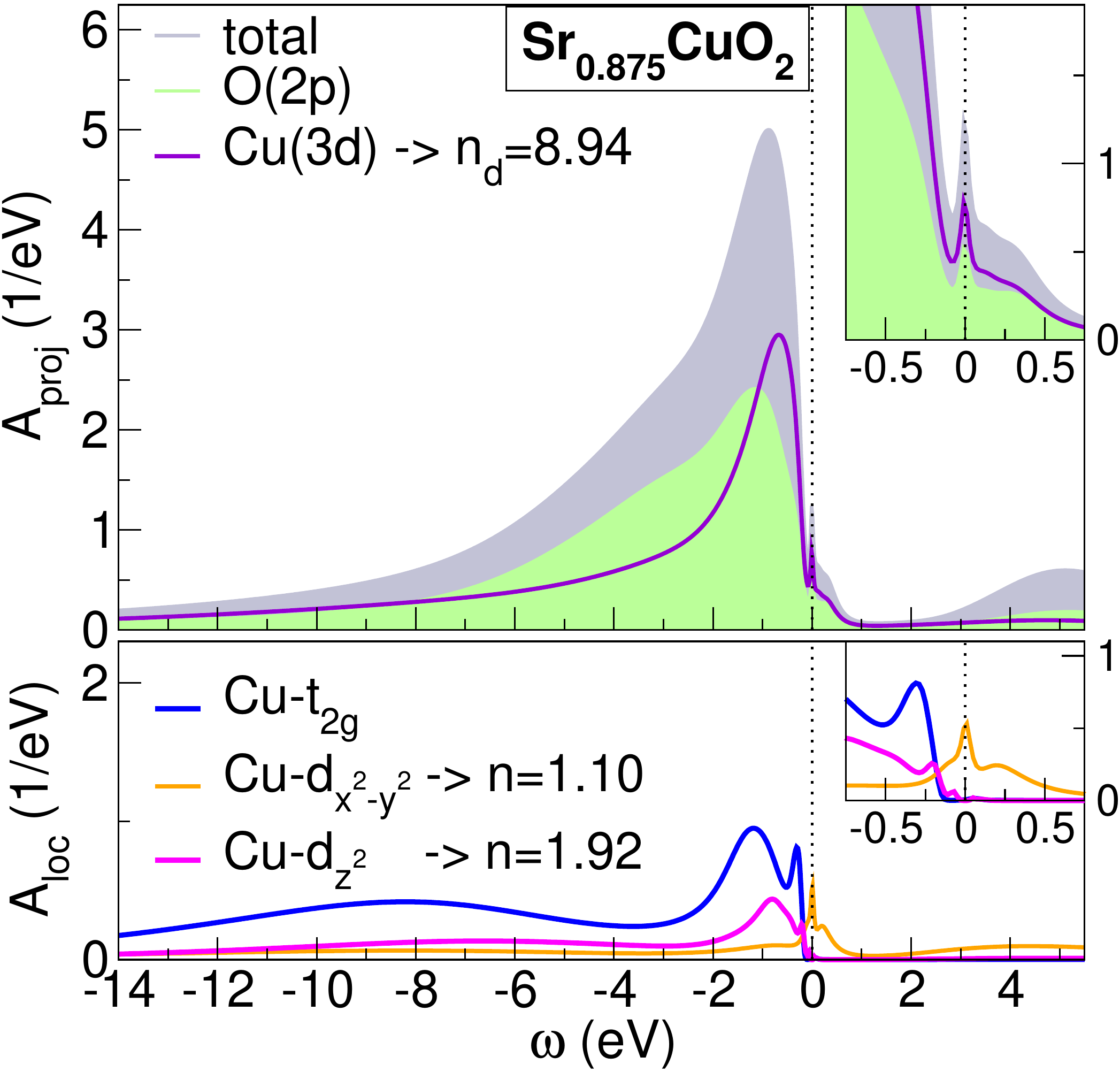}\\[-0.2cm]
\raggedright(a)\hspace*{5.6cm}(b)\hspace*{5.6cm}(c)
\caption{(color online) DFT+sicDMFT spectral data of hole-doped materials. Upper parts:
total as well as site- and orbital-projected spectral function (inset: low-energy blow up),
with filling $n_d$ of the TM$(3d)$ shell.
Lower parts: TM$(3d)$ local spectral function, with filling $n$ of the $e_g$ orbitals.
(a) Nd$_{0.875}$Sr$_{0.125}$NiO$_2$, (b) Nd$_{0.750}$Sr$_{0.250}$NiO$_2$ and
(c) Sr$_{0.875}$CuO$_2$.}
\label{fig:dop}
\end{figure*}
Let us turn to the electronic states with hole doping. The doped cases are realized
by $2\times 2\times 2$ supercells of the nickelate and cuprate unit cells, respectively.
This amounts to eight TM sites in the primitive cell, compared to only one in the 
stoichiometric unit cell. In the nickelate, Nd atoms are replaced by Sr atoms~\cite{li19} 
by a fraction $x$, while in the cuprate, Sr vacancies~\cite{ter96} are generated by a fraction 
$y$. For Nd$_{1-x}$Sr$_x$NiO$_2$, two supercells are constructed: replacing on Nd atom
$(x=0.125)$, and replacing two Nd atoms $(x=0.250)$. For Sr$_y$CuO$_2$, one vacant 
Sr site is introduced $(y=0.125)$. Structural relaxation of the atomic positions within DFT+U
leads to a minor modulation of mainly the TM-O bond lengths, up to 0.5\% of the respective
stoichiometric distances. Notably, since each Sr vacancy gives rise to two holes,
the actual doping level $\delta$ amounts to $\delta=x=2y$. Note also again that we are
here interested in principle effects and differences, and a very thorough many-body study
of late quasi-twodimensional TM oxides at finite doping will ask for correlation effects 
beyond single-site DMFT (see e.g. Refs.~\onlinecite{mai05,roh18} for reviews). 

Figure~\ref{fig:dop} displays the spectral properties of the three hole-doping scenarios.
In all cases, the electronic spectrum is shifted upwards in energy with doping, in line
with available experimental data for IL cuprates~\cite{ter96,che91}.
The doped charge-transfer insulator shows in Fig.~\ref{fig:dop}c the spectral 
signature expected from a hole-doped cuprate~\cite{che91}: filling of the gap and a 
resonance at $\varepsilon^{\hfill}_{\rm F}$, both with a substantial contribution
from O$(2p)$ states. 
Comparing the occupation numbers, $\delta_{\rm Cu}=n_d^{\rm undoped}-n_d^{\rm doped}$=0.10 
holes per site are located on copper, while $\delta_{\rm O}=\delta-\delta_{\rm Cu}$=0.15 holes 
per site are attributed to oxygen. The low-energy resonance marks the itinerant 
Zhang-Rice singlet physics~\cite{zha88}. Hence, on the local Cu$(3d)$ level, only the 
$x^2-y^2$ orbital is contributing to the low-energy resonance. The situation is more 
intriguing for Nd$_{1-x}$Sr$_x$NiO$_2$ (cf. Fig.~\ref{fig:dop}a,b). First, significant 
low-energy Zhang-Rice physics cannot readily be identified, the spectrum close to 
$\varepsilon^{\hfill}_{\rm F}$ is strongly Ni$(3d)$ dominated. Comparing the occupation 
numbers as before, a much larger relative hole count of $\delta_{\rm Ni}=0.09$ for 
$x=0.125$ and of $\delta_{\rm Ni}=0.22$ for $x=0.250$ on the TM site is obtained. 
Second, while for $x=0.125$ the $x^2-y^2$ orbital is mainly susceptible to doping, for 
$x=0.250$ the $z^2$ orbital joins in~\cite{wer20,jia19} and takes over. Because of the missing 
Zhang-Rice physics, holes are not easily shared between Ni and O sites. Thus for larger 
hole content, the Ni site has to provide access to additional orbital degrees of freedom.
This would result in a $x^2-y^2$ vs. $z^2$ multi-orbital competition at the experimental 
$x=0.2$ scenario for superconductivity~\cite{li19}. The prominent $z^2$ role is also
supported by the favorable hybridizations established at stoichiometry. 
Note that the self-doping band itself is depleted already for the given dopings. While
some minor Nd spectral content may still be observed at low energy, the actual electron 
pockets are shifted upwards by $\sim 0.9\,$eV for $x=0.125$.

\textit{Summary and discussion.---}
There are key differences between the infinite-layer TM oxides NdNiO$_2$ and SrCuO$_2$.
The cuprate is a charge-transfer insulator at strong coupling, whereas the nickelate
remains a non-insulating, self-doped system at large interaction strength. Still,
in both system the TM-$d_{x^2-y^2}$ state is half-filled localized (or very close to
such a regime). In the IL architecture it is naturally expected that the missing
apical oxygens allow for novel/modified hybridization scenarios for the TM$(3d)$ orbitals
with the surrounding. In the cuprate case, this does not lead to major qualitative 
changes compared to the perovskite case, since hybridization with non-oxygen orbitals 
leads to states too high in energy to effect low-energy properties. 
On the other hand for rare-earth nickelates, the $5d$ orbitals of the rare-earth ion enable 
low-energy hybridization with Ni$(3d)$, i.e. additional band crossing at 
$\varepsilon^{\hfill}_{\rm F}$ from self doping. 
The Fermi surface becomes multi-sheeted, but still encompassing one electron. At strong 
coupling, where experiment seemingly places the material~\cite{li19}, the $d_{x^2-y^2}$ 
dominated sheet becomes gapped, but Luttinger theorem ensures the survival of the 
self-doping sheet.
 
For hole doping, the role of the self-doping band becomes minor for sizable dopings. 
However, a strong TM$(3d)$-O$(2p)$ coupling at low energy is absent for the nickelate, and 
furthermore, $x^2-y^2$ and $z^2$ compete close to the Fermi level. On the other hand, 
hole-doped SrCuO$_2$ displays the expected $x^2-y^2$ dominance even for the present 
overdoped scenario in view of superconductivity. One may speculate if the reduced 
$T_{\rm c}$ for the nickelate compared to cuprates may be the result of that competition.
For instance, due to loss of coherence in the interplay between two rather 
differently characterized orbital settings. The effective 
single-TM$(3d)$ orbital together with the strong Zhang-Rice physics would then be
key to the high $T_{\rm c}$ of cuprates.

Further studies on IL nickelates and akin systems are highly desirable to fathom the
designing options in view of raising $T_{\rm c}$~\cite{nom19}. In this context, the route 
along the metal-insulator transition in the oxygen-deficient LaNiO$_{3-x}$ perovskite for 
$x\geq 0.25$ with apparent oxygen-vacancy ordering is also 
noteworthy~\cite{say94,san96,wan18}.

\begin{acknowledgments}
The author thanks A. J. Millis and R. Valent{\'i} for helpful discussions.
Financial support from the German Science Foundation (DFG) via the project LE-2446/4-1 
is gratefully acknowledged. 
Computations were performed at the University of Hamburg and the JUWELS 
Cluster of the J\"ulich Supercomputing Centre (JSC) under project number hhh08.
\end{acknowledgments}

\bibliographystyle{apsrev}
\bibliography{bibinf}

\end{document}